\documentstyle[art12,bezier,emlines2]{article}
\normalbaselines
\begin{document}
\voffset=-3cm
\begin{center}
{\Large\bf 
Holography and the origin of anomalies
}
\\[5mm]
{\bf B. P. Kosyakov}\\[5mm]
{\it Russian Federal Nuclear Center -- VNIIEF, 
Sarov 607190, Russia}
\end{center}
\begin{abstract}
The holographic principle is represented as the well-known de Alfaro, Fubini 
and Furlan correspondence between the generating functional for 
the Green functions of the Euclidean quantum field theory in $D$ dimensions and
the Gibbs average for the classical statistical mechanics in $D+1$ dimensions.
This  correspondence is used to explain the origin of quantum anomalies
and the irreversibility in the classical theory.
The holographic mapping of a classical string field theory 
onto a local quantum field theory is outlined.
\end{abstract}

\section{Introduction}\label{Intr}
Anomalies play an important role in high energy physics and
string theory \cite{Jackiw}. 
Technical aspects of this phenomenon have received much 
attention \cite{Treiman}.
Meanwhile the origin of anomalies still remains an enigma.
The aim of this paper is to show that the holographic principle can clarify 
this issue.

The holographic principle was first proclaimed by 't Hooft \cite{Hooft} and 
Susskind \cite{Susskind} in the context of black holes. 
According to this principle, information on degrees of freedom 
inside a volume can be projected onto a surface (also called the screen) which 
encloses this volume.
For lack of the generally accepted formulation, we will turn to provisional 
versions of this principle discussed in the 
literature.
Their essence is as follows: A classical theory (which includes gravity), 
describing phenomena within a volume, can be formulated as a quantum theory
of these phenomena (without gravity) projected onto the boundary of this 
volume.  
In such a form, the holographic principle was confirmed in \cite{Maldacena} 
where the consistency between semiclassical supergravity 
in an anti-de Sitter space and quantum superconformal Yang-Mills theory on
the boundary of this space was revealed. 

There is reason to believe that the holographic principle is valid also when 
gravity is excluded.
Then it can be realized through a remarkable 
de Alfaro, Fubini and Furlan dualism \cite{Alfaro}.
They argued that the generating functional for Green's functions of the
Euclidean quantum field theory in $D$ dimensions coincides with the Gibbs 
average for the classical statistical mechanics in $D+1$ dimensions.
In other words, there exists a correspondence (the AFF dualism) between 
the classical picture in a spacetime ${M}_{D+1}$ and the quantum picture 
in $D$-dimensional sections of ${ M}_{D+1}$ at any fixed instants.
Such sections play the role of $D$-dimensional screens carrying the hologram 
of what happens in ${M}_{D+1}$. 

We recall the idea of the AFF dualism
by the example of a system described by scalar field $\phi(x)$. 
Let the system be located in a $D$-dimensional Euclidean spacetime and 
specified by the Lagrangian ${\cal L}$.
One introduces a fictious time $t$\footnote{There is a similar much-studied 
scheme of quantization discovered by Parisi and Wu, Sci. Sinica {\bf 24} (1981)
483, the stochastic quantization (for a review, 
see \cite{Namiki}), where the fictious time is also used.
The formal equivalence of the de Alfaro, Fubini and Furlan method to that of
Parisi and Wu was established in \cite{Gozzi}. 
Notice that the holographic principle can also be realized via the 
stochastic quantization, \cite{vandeBruck}, but this approach appears to
be less helpful in understanding the origin of anomalies.}.
The field becomes a function of the Euclidean coordinates $x_1,\ldots,x_{D}$ 
and fictious time $t$, $\phi=\phi(x,t)$.
If ${\scriptstyle\frac12}(\partial\phi/\partial t)^2$ is treated as the
kinetic term, and $\cal L$ the potential energy term, then one defines a new
Lagrangian 
\[{\tilde{\cal L}}={\scriptstyle\frac12}(\partial\phi/\partial t)^2-{\cal L}\]
generating the evolution in $t$.
The associated Hamiltonian is 
\[{\tilde{\cal H}}={\scriptstyle\frac12}\,\pi^2+{\cal L}\] 
where $\pi=\partial{\tilde{\cal L}}/\partial{\dot\phi}=
\partial\phi/\partial t$ stands for the conjugate momentum which is assumed to
obey the classical  Poisson bracket 
\begin{equation}
\{\phi(x,t),\pi(y,t)\}=\delta^D(x-y).
\label
{P-B}
\end{equation}                                            
It is easy to see that the Gibbs average for an ensemble with the temperature 
$kT=\hbar$
\begin{equation}
{\cal Z}
[J]=\int {\cal D}\pi {\cal D}\phi\,\exp\Bigl(-\frac{1}{kT}
\int d^Dx\,({\tilde{\cal H}}+J\phi)\Bigr)
\label
{Z-cl}
\end{equation}                                            
turns to the generating functional for the quantum Green functions 
\begin{equation}
Z[J]=\int {\cal D}\phi\,\exp\Bigl(-\frac{1}{\hbar}\int d^Dx\,({\cal L}+J\phi)
\Bigr)
\label
{Z-q}
\end{equation}                                            
upon taking the Gaussian integral over $\pi$.
Note that the holographic mapping of the bulk picture in ${M}_{D+1}$ 
onto the screen picture in a section of ${ M}_{D+1}$ at any instant $t$ is 
ensured by the Liouville theorem.
Indeed, although $\phi(x,t)$ and $\pi(x,t)$ evolve in $t$, the elementary 
volume 
in phase space ${\cal D}\pi{\cal D}\phi$, and with it the Gibbs average 
${\cal Z}$ are  $t$-independent. 

Thus, it is meaningless to ask whether a given realm is classical or quantum.
It may appear both as classical and quantum, but these two looks
pertain to spacetimes of nearby dimensions\footnote{By comparison,
't Hooft in his recent work  \cite{Hooft99} claimed: ``In our theory, quantum 
states are not the primary degrees of freedom.
The primary degrees of freedom are deterministic states.''}.
To identify the realm, one should only indicate $D$.
For example, assigning $D=4$ to the electromagnetic realm, we bear in mind
that it can be grasped by either some `quantum' 4D Lagrangian ${{\cal L}}$ 
or the associated `classical' 5D Lagrangian ${\tilde{\cal L}}$.

It is clear that the conventional procedure of quantization only shifts
the seat to another realm: In lieu of the initial classical system living in
$D$ dimensions, a new classical system living in $D+1$ dimensions emerges.
As is well known, symmetries of classical Lagrangians may be sensitive to the 
dimension; some of them are feasible only for a single $D^\star$, 
while another only for $D=2n$.
On the other hand, given a quantized $D^\star$-dimensional theory, we deal
actually with the holographic image of classical $D^\star+1$-dimensional 
theory, and the symmetry under examination is sure to be missing from it.
Therein lies the reason for the symmetry damage due to quantization known as
the `anomaly'.
The responsibility for this damage rests with the fact that 
the compared classical descriptions merely differ in dimensions.
Such an interpretation is exemplified by the conformal (Weyl) anomaly 
\footnote{The anomaly of this kind was recently analyzed in the holographic 
context in \cite{HenningsonSkenderis}.} in Sec. \ref{anomalies}. 

In the general case, the AFF dualism does not imply that the classical 
$D+1$-dimensional theory is equivalent to its quantum $D$-dimensional 
counterpart.
There are two reasons for the lack of equivalence.
First, the field theory can suffer from ultraviolet divergences, and the
equality ${\cal Z}=Z$, strictly speaking, does not have mathematical sense. 
The dissimilarity of classical divergences from quantum ones keeps
the expressions ${\cal Z}$ and $Z$ from being equal.
This lack of equivalence  prohibits derivation of exact expressions for 
anomalies by a direct comparison of two classical actions of adjacent 
dimensions. 
Furthermore, it gives rise to a {\it classical anomaly} related to the 
violation of reversibility on the classical level with preserving this
symmetry on the quantum level.
Section \ref{irreversib} is devoted to this phenomenon.

Second, with a one-to-one mapping of a classical picture onto a quantum
one, the image would display a violation of the classical
determinism implying the loss of information on seemingly unique solutions
of the Cauchy problems for classical dynamical equations (or,
alternatively, the inverse image would
display a piece of the quantum coherence).   
Nevertheless, such is the case if classical string field theory 
is mapped onto local quantum field theory, i. e., for $D=1$.
This issue is sketched in Sec. \ref{special}, and will be detailed elsewhere.

\section{Conformal anomaly}\label
{anomalies}
The holography is particularly suited for studying the origin of 
anomalies since one may repeatedly handle classical actions alone.
Canonical transformations leave the Poisson bracket (\ref{P-B}) invariant,
hence the measure of integration in (\ref{Z-cl}) ${\cal D}\pi{\cal D}\phi$ 
has a corresponding canonical invariance property enabling one to perform
canonical transformations without having to introduce a Jacobian. 

Let us consider the conformal (Weyl) transformation of the metric 
\begin{equation}
g_{\mu\nu}\to e^{2\varepsilon}\,g_{\mu\nu}
\label
{conf-transf}
\end{equation}                                           
resulting in the associated Noether current, the energy-momentum tensor
\begin{equation}
\Theta^{\mu\nu}=\frac{2}{\sqrt{- g}}\,\frac{\delta}{\delta g_{\mu\nu}}\,\sqrt{- g}\,{\cal L}.
\label
{Theta-def}
\end{equation}                                           
The conformal invariance of the classical action $S$ is attained if
\begin{equation}
\delta S=2\varepsilon\,g_{\mu\nu}\Theta^{\mu\nu}=0.
\label
{Theta-mu-mu}
\end{equation}                                           
It follows from (\ref{Theta-def}) that, in the classical $D+1$-dimensional 
Yang-Mills theory with 
\[
{\cal L}_{\rm YM}=-\frac{1}{4\Omega_{D-1}\alpha}\,{\rm tr}\,F_{\alpha\beta}
F^{\alpha\beta}, 
\]
$\Theta^{\mu\nu}$ is written as
\begin{equation}
\Theta^{\mu\nu}=\frac{1}{\Omega_{D-1}\alpha}\,{\rm tr}\,(F^{\mu\alpha}
F_{\alpha}^{\hskip1.5mm\nu}+
{\frac{1}{4}}\,\eta^{\mu\nu}F_{\alpha\beta}F^{\alpha\beta}).
\label
{Theta}
\end{equation}                                           
Here, $\Omega_{D-1}$ is the area of a $D-1$-dimensional unite sphere, and 
$\alpha$ the Yang-Mills coupling.
The expression (\ref{Theta}) makes it clear that the condition 
(\ref{Theta-mu-mu}) is fulfilled only for $D+1=4$.
Yang-Mills equations are conformally invariant only in a four-dimensional 
spacetime.

The quantization of the classical 4D Yang-Mills theory 
culminates in the classical 5D Yang-Mills theory where one finds 
$\Theta^{\mu}_{\hskip1mm\mu}\ne 0$. 
This is how the conformal anomaly occurs.

Technically, the conformal invariance breakdown in quantum field theory
is traced to the mass scale $\mu$ introduced for the normalization of $\alpha$.
Coleman and Weinberg \cite{Coleman} called this the
`dimensional transmutation'.
Such a transmutation is missing from the quantum 3D Yang-Mills theory, thereby 
the conformal invariance of its AFF dual, the classical 4D Yang-Mills theory, 
is ensured. 
Why does this happen?
In the next section, we will see that the quantum 3D theory is 
super-renormalizable, the vacuum polarization is weak, there is no infinite 
charge renormalization, and the need for $\mu$ disappears.

As to the quantum 4D theory, vacuum polarization effects are significant here, 
and $\alpha$ becomes the running coupling constant  $\alpha(q^2/\mu^2)$,
implying that 
\begin{equation}
\Theta^{\mu}_{\hskip1mm\mu}=\frac{2g^{\mu\nu}}{\alpha^2}\,\frac{\partial
\alpha}{\partial g^{\mu\nu}}\,\frac{1}{16\pi}\,{\rm tr}\,F^2=
\frac{1}{8\pi}\,\frac{\beta(\alpha)}{\alpha^2}\,{\rm tr}\,F^2
\label
{Theta-mu-mu-q}
\end{equation}                                           
where $\beta=\partial\alpha/\partial\log q^2$ is the Gell-Mann-Low function. 
We encounter the $q^2$-dependent expression as opposed to the naive expression
\[
\Theta^{\mu}_{\hskip1mm\mu}=
\frac{1}{8\pi^2\alpha}\,{\rm tr}\,F^2
\]
which might be derived from the classical 5D Yang-Mills action.
In the classical picture, there are no creations and annihilations of 
particles, the vacuum polarization is absent, and $\alpha$ is not renormalized.
We will establish further, that in the classical 5D theory, 
not all ultraviolet divergences can be absorbed in the mass renormalization.
For some of them to be absorbed, the presence of higher derivative terms in
the Lagrangian is necessary; a consistent description is achieved only
within the scope of a rigid dynamics.
Thus the AFF dualism in the Yang-Mills theory turns out to be ruined for 
$D\ge 4$.

The action for the bosonic string in a flat target space, 
describing two-dimensional gravity on the world sheet of the string,
\[
S=\frac{1}{4\pi}\int d^2\sigma\,\sqrt{-g}\,\Biggl(\frac{1}{\alpha'}\,
g^{ab}\partial_a X^\mu\partial_b X_\mu- \frac{1}{k}\,R\Biggr),
\]
is invariant under the Weyl transformations (\ref{conf-transf}).
The variation of the action with respect to the metric yields the 
energy-momentum tensor
\begin{equation}
T_{ab}=\frac{1}{4\pi\alpha'}\,(\partial_a X^\mu\partial_b X_\mu
-{\frac12}\,g_{ab}\,\partial_c X^\mu\partial^c X_\mu)-
\frac{1}{4\pi k}\,(R_{ab}-{\frac12}\,g_{ab}\,R).
\label
{T-a-b}
\end{equation}                                           
It follows that classical two-dimensional gravity is conformally
invariant, while this invariance disappears in classical three-dimensional 
gravity.
Nevertheless, $T^{a}_{\hskip1mm a}$ fails to be obtained by mere contraction 
of Eq. (\ref{T-a-b}) in the latter case.
The vacuum polarization is again significant, so that
$T^{a}_{\hskip1mm a}\propto\beta(\alpha')$. 
As is weell-known, for a flat $D$-dimensional target space,
\[
\beta=\frac{D-26}{12}.
\]
Thus, for a purely gravitational two-dimensional realm, we have only 
a formal AFF correspondence, not equivalence.

\section{Irreversibility}\label
{irreversib}
The violation of the AFF equivalence by ultraviolet divergences
can be regarded as a kind of anomalies.
As an example let us consider the occurrence of irreversibility in a classical 
picture, with reversibility in the dual quantum picture being left intact.
To be specific, take a $D$-dimensional quantum realm described 
by the 
scalar electrodynamics Lagrangian
\begin{equation}
{\cal L}_0=(\partial^\mu+ig_0 A^\mu)\phi\,(\partial_\mu-ig_0 A_\mu){\bar
{\phi}}-m_0^2\phi\,{\bar{\phi}}-\frac{\lambda_0}{4}(\phi\,{\bar{\phi}})^2
-\frac{1}{4\Omega_{D-2}}\,F_{\mu\nu}F^{\mu\nu}.
\label
{L-scal}
\end{equation}                                            

In the dual $D+1$-dimensional classical realm, the scalar field $\phi$ can be
treated as a Lagrangian coordinate of a continuous medium that evolves
in time $t=x_{D+1}$.
However, such a model is inconvenient for analysis of ultraviolet properties,
and we will discuss its discrete analogue, the system with very large number
(strictly speaking, $\infty^D$) of charged point particles.
Assume that the action for the classical particle interacting with 
electromagnetic field is of the usual form 
\begin{equation}
S=-\int d\tau\,(m_0\sqrt{v\cdot v}+ g\,v\cdot A)
\label
{action}
\end{equation}                                            
where $v^\mu\equiv{\dot z}^\mu\equiv dz^\mu/d\tau$ is the
$D+1$-velocity of the particle, and $\tau$ the proper time.

Features of dual AFF pairs corresponding to several values of $D$ are
summarized in Table 1.
Let us begin with the line `Ultraviolet behavior' indicating the dependence
of observables on the cutoff $\Lambda$.
The maximal power of  $\Lambda$ increases with $D$ to give the 
progressive violation of the AFF dualism up to its complete failure
above $D=4$. 
\begin{center}
\begin{tabular}{|c|c||c|c||c|c||c|c|}  
\multicolumn{8}{c}{TABLE 1.\quad{\bf  Dualities in scalar electrodynamics}}\\[2mm]\hline 
\multicolumn{8}{|c|}{ {\it Spacetime dimension of the quantum picture
}}\\ \hline
\multicolumn{2}{|c||}{$D=1$} & \multicolumn{2}{|c||}{$D=2$} & \multicolumn{2}{|c||}{$D=3$}&\multicolumn{2}{|c|}{$D=4$}   \\ \hline\hline
\multicolumn{8}{|c|}{
\it Dual AFF pairs}\\ \hline
\multicolumn{2}{|c||}{1D$_{\rm quant}$ -- 2D$_{\rm class}$} & \multicolumn{2}{|c||}{2D$_{\rm quant}$ -- 3D$_{\rm class}$} & \multicolumn{2}{|c||}{3D$_{\rm quant}$ -- 4D$_{\rm class}$}&\multicolumn{2}{|c|}{4D$_{\rm quant}$ -- 5D$_{\rm class}$}\\ \hline\hline
\multicolumn{8}{|c|}{
\it Ultraviolet behavior}\\ \hline
$\Lambda^0$ & $\Lambda^0$ & $\log\Lambda$ & $\log\Lambda $  & $\Lambda$ & $\Lambda$ & $\Lambda^2$, $\log\Lambda$ & $\Lambda^2$, $\log\Lambda$ \\ 
\hline\hline
\multicolumn{8}{|c|}{
\it Renormalizability}\\ \hline
{\small  finite} &{\small  finite}  & {\small super-}    & {\small ren'zable}  & {\small super-}    & {\small ren'zable} & {\small ren'zable} & {\small non-} \\ 
{\small        } &{\small        }  & {\small ren'zable} & {\small }         & {\small ren'zable} & {\small          } &{\small           } & {\small ren'zable}\\ \hline\hline
\multicolumn{8}{|c|}{
\it Reversibility}\\ 
\hline
{\small holds}     & {\small holds}     & {\small holds}     & {\small holds    } & {\small holds   }   & {\small weak}  & {\small holds}& {\small topolog}\\
                 &                  &                  &                  &                   & {\small violat}  &             &  {\small violat} \\\hline
\end{tabular}
\end{center}

Among $D+1$-dimensional classical quantities, the dependence on $\Lambda$ is 
inherent in the energy-momentum\footnote{The angular momentum has the same 
ultraviolet behavior}
of electromagnetic field generated by a point particle, 
\begin{equation}
P_\mu=\int  d\sigma^\nu\,\Theta_{\mu\nu},
\label
{P-mu}
\end{equation}                                            
where the integration is performed over a $D$-dimensional spacelike 
hypersurface.
In the static case when electromagnetic field can be specified by the
potential $\varphi$ satisfying the Poisson equation
\begin{equation}
\Delta\varphi ({\bf x})=-\Omega_{D-1}\,\rho({\bf x})
\label
{Poisson}
\end{equation}                                            
with 
\begin{equation}
\rho({\bf x})=g\,\delta^{D}({\bf x}),
\label
{rho}
\end{equation}                                            
one has
\begin{equation}
\varphi({\bf x})=g
\cases{\vert\,{\bf x}\vert^{2-D},  & $D\ne 2$,\cr
\log\vert\,{\bf x}\vert, & $D=2$.\cr}
\label
{varphi}
\end{equation}                                            
From (\ref{rho}) and (\ref{varphi}) in combination with the static 
expression for the particle self-energy 
\begin{equation}
\delta m={\frac12}\int d^{D}{\bf x}\,\rho({\bf x})\,\varphi({\bf x})\,
={\frac12}\,g^2\varphi(0),
\label
{delta m}
\end{equation}                                            
one obtains leading $\Lambda$-dependences indicated in Table 1.

For small deviations from statics, Eq. (\ref{P-mu}) acquires 
the form 
\begin{equation}
P_\mu=c_1\,v_\mu+ c_2\,{\dot v}_\mu +c_3\,{\ddot v}_\mu +\ldots
\label
{quasi}
\end{equation}                                            
where the variables ${v}_\mu$, ${\dot v}_\mu$, etc., relate to the point source
of the field at a given instant.
It is clear that $c_1=\delta m$.
From dimensional considerations, one finds also that 
$c_i/c_{i+1}\sim\Lambda$.

In the 5D$_{\rm class}$ case, the integration of Eq. (\ref{P-mu}) 
is carried out over a four-dimensional hypersurface, and, therefore,
only even powers of $\Lambda$ are nonzero:
\begin{equation}
c_1\sim\Lambda^2,\quad c_2=0,\quad c_3\sim\log\Lambda.
\label
{c-i}
\end{equation}                                            
The $\Lambda^2$ term is absorbed by the mass renormalization, but there is 
no parameter in the action (\ref{action}) suitable for the absorption 
of the logarithmic divergence, and the 5D$_{\rm class}$ theory turns out to be 
{\it nonrenormalizable}.
To remedy the situation, to the action (\ref{action}) must be added a term 
with higher derivatives \cite{k9} of the type
\begin{equation}
-\kappa_0 \int  d\tau\,
\frac{1}{\sqrt{{v}\cdot{v}}}\,\Biggl(\frac{d}{d\tau}\frac{{v}^\mu}
{\sqrt{{v}\cdot{v}}}\Biggr)^2.
\label{222}
\end{equation}                                       

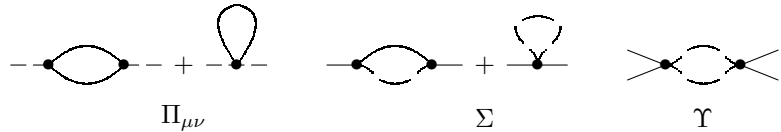
\begin{figure}[htb]
\begin{center}
\unitlength=1.00mm
\special{em:linewidth 0.4pt}
\linethickness{0.4pt}
\begin{picture}(112.00,18.00)
\emline{10.00}{10.00}{1}{12.00}{10.00}{2}
\emline{13.00}{10.00}{3}{15.00}{10.00}{4}
\emline{25.00}{10.00}{5}{27.00}{10.00}{6}
\emline{28.00}{10.00}{7}{30.00}{10.00}{8}
\emline{36.00}{10.00}{9}{38.00}{10.00}{10}
\emline{39.00}{10.00}{11}{41.00}{10.00}{12}
\emline{42.00}{10.00}{13}{44.00}{10.00}{14}
\emline{52.00}{10.00}{15}{56.00}{10.00}{16}
\emline{66.00}{10.00}{17}{70.00}{10.00}{18}
\emline{76.00}{10.00}{19}{84.00}{10.00}{20}
\emline{92.00}{12.00}{21}{97.00}{10.00}{22}
\emline{97.00}{10.00}{23}{92.00}{8.00}{24}
\emline{107.00}{10.00}{25}{112.00}{12.00}{26}
\emline{112.00}{8.00}{27}{107.00}{10.00}{28}
\bezier{56}(15.00,10.00)(20.00,15.00)(25.00,10.00)
\bezier{56}(15.00,10.00)(20.00,5.00)(25.00,10.00)
\bezier{56}(40.00,10.00)(35.00,17.00)(40.00,18.00)
\bezier{56}(40.00,10.00)(45.00,17.00)(40.00,18.00)
\bezier{56}(56.00,10.00)(61.00,15.00)(66.00,10.00)
\bezier{16}(59.00,8.00)(61.00,7.00)(63.00,8.00)
\bezier{16}(78.00,16.00)(80.00,17.00)(82.00,16.00)
\bezier{12}(77.00,15.00)(77.00,14.00)(78.00,13.00)
\bezier{12}(83.00,15.00)(83.00,14.00)(82.00,13.00)
\bezier{16}(100.00,12.00)(102.00,13.00)(104.00,12.00)
\bezier{16}(100.00,8.00)(102.00,7.00)(104.00,8.00)
\put(15.00,10.00){\makebox(0,0)[cc]{$\bullet$}}
\put(25.00,10.00){\makebox(0,0)[cc]{$\bullet$}}
\put(40.00,10.00){\makebox(0,0)[cc]{$\bullet$}}
\put(33.00,10.00){\makebox(0,0)[cc]{$+$}}
\put(33.00,3.00){\makebox(0,0)[cc]{$\Pi_{\mu\nu}$}}
\put(56.00,10.00){\makebox(0,0)[cc]{$\bullet$}}
\put(66.00,10.00){\makebox(0,0)[cc]{$\bullet$}}
\put(73.00,10.00){\makebox(0,0)[cc]{$+$}}
\put(80.00,10.00){\makebox(0,0)[cc]{$\bullet$}}
\put(73.00,3.00){\makebox(0,0)[cc]{$\Sigma$}}
\put(97.00,10.00){\makebox(0,0)[cc]{$\bullet$}}
\put(107.00,10.00){\makebox(0,0)[cc]{$\bullet$}}
\put(102.00,3.00){\makebox(0,0)[cc]{$\Upsilon$}}
\bezier{8}(80.00,10.00)(80.00,11.00)(79.00,12.00)
\bezier{8}(80.00,10.00)(80.00,11.00)(81.00,12.00)
\bezier{8}(97.00,10.00)(98.00,10.00)(99.00,11.00)
\bezier{8}(97.00,10.00)(98.00,10.00)(99.00,9.00)
\bezier{8}(107.00,10.00)(106.00,10.00)(105.00,11.00)
\bezier{8}(107.00,10.00)(106.00,10.00)(105.00,9.00)
\bezier{8}(56.00,10.00)(57.00,10.00)(58.00,9.00)
\bezier{8}(66.00,10.00)(65.00,10.00)(64.00,9.00)
\end{picture}
\caption{One-loop diagrams of the scalar electrodynamics}
\label
{diagram}
\end{center}
\end{figure}

As to the dual quantum quantities, we are directly concerned with
the polarization operator $\Pi_{\mu\nu}$, the scalar self-energy $\Sigma$, and 
the scalar self-interaction $\Upsilon$.
The relevant one-loop diagrams are depicted in Fig. \ref{diagram}.
(The light-light scattering revealing a more soft ultraviolet behavior is 
immaterial here.)
Using a gauge-invariant regularization, $\Pi_{\mu\nu}$ can be cast as 
$\Pi_{\mu\nu}(q)=(q^2\eta_{\mu\nu}-q_\mu q_\nu)\,\Pi(q^2)$.
An elementary Feynman technique leads to
\[
\Sigma\sim\int\frac{d^Dk}{k^2},
\qquad
\Pi\sim\int\frac{d^Dk}{(k^2)^2},
\qquad
\Upsilon\sim\int\frac{d^Dk}{(k^2)^2}.
\]
It follows
\begin{equation}
\Sigma\sim
\cases{\Lambda^{D-2},  & $D\ne 2$,\cr
\log\Lambda, & $D=2$,\cr}
\qquad
\Pi\sim
\cases{\Lambda^{D-4},  & $D\ne 4$,\cr
\log\Lambda, & $D=4$,\cr}
\qquad
\Upsilon\sim
\cases{\Lambda^{D-4},  & $D\ne 4$,\cr
\log\Lambda, & $D=4$.\cr}
\label
{UV}
\end{equation}                                            

The comparison of (\ref{varphi})--(\ref{delta m}) and (\ref{c-i}) with 
(\ref{UV}) shows that the divergence powers in the 
5D$_{\rm class}$ theory 
are the same as those in the 
4D$_{\rm quant}$ theory.
Nevertheless, the latter is renormalizable \cite{Salam}: All the primitively 
divergent diagrams refer to the quantities
$\Pi$, $\Sigma$ and $\Upsilon$ which renormalize, respectively, 
$g_0$, $m_0$ and $\lambda_0$.

Since the bar Lagrangian ${\cal L}_0$ and counter-terms in the 4D$_{\rm 
quant}$ theory have the same structure, and  ${\cal L}_0$ is invariant 
under time reversal $t\to -t$, the renormalized quantum dynamics is reversible.
Meanwhile, in the dual 5D$_{\rm class}$ theory, 
the potential energy of two charged particles $U(r)=-g^2/r^{2}$ is more 
singular than the centrifugal term $J/r$, and the fall to the centre is 
inevitable unless the action is modified {\it ad hoc} by the addition of
terms with higher derivatives.
The fall to the centre is evidence of irreversibility.
Note that we encounter a topologically deficient picture: The mere presence of
a certain variant of evolution (the fall to the centre) rules out the
possibility of the inverse variant of evolution (splitting and subsequent
departure of the merged particles).
The violation of the AFF dualism is due to the fact that the 5D$_{\rm class}$ 
picture is devoid of the vacuum polarization which is responsible for 
infinite renormalization of $g_0$ and $\lambda_0$.

Let us turn to the case $D=3$.
Now $\Pi$ and $\Upsilon$ are finite, hence the coupling constants 
are not subject to infinite renormalization.
The divergences of both $\Sigma$ and $\delta m$ are absorbed by the mass
renormalization.
The number of parameters contained in the Lagrangians are sufficient for the 
removal of all divergences, so, at first glance, the AFF equivalence is the 
case. 
But this is wrong.
The mass renormalization makes a finite {\it mark} which is more deep in the
classical picture than in the quantum one.
The renormalized quantum dynamics is reversible.
By contrast, in the renormalized classical dynamics, reversibility is violated.
Indeed, the coefficients $c_i$ in (\ref{quasi}) are
\cite{teit}
\[c_1\sim \Lambda,\quad 
c_2=-{\frac23}\,g^2,\quad c_3\sim \Lambda^{-1},\]
whence it follows the expression for the four-momentum of the 
`dressed' particle 
\begin{equation}
p_\mu=m\,v_\mu-{\frac23}\,g^2\,{\dot v}_\mu,
\label
{p-mu}
\end{equation}                                            
($m$ is the renormalized mass), and the equation of motion of the
`dressed' particle 
\begin{equation}
m\,{\dot v}^\mu-{\frac23}\,g^2\,({\ddot v}^\mu+ {\dot v}^2 v^\mu)=
f^\mu.
\label
{LD}
\end{equation}                                            
This is the Lorentz-Dirac equation which is known to be not invariant under time reversal 
$\tau\to -\tau$.
This irreversibility is due to the energy dissipation stemming from 
radiation of electromagnetic waves.

Many attempts to derive Eq. (\ref{LD}) from some Lagrangian have not met with
success.
Thus the partition function (\ref{Z-cl}) whose construction is based on the 
Hamiltonian (which in turn implies the availability of the corresponding 
Lagrangian) has nothing to do with the renormalized classical dynamics.
This violates the AFF dualism.

The pecularity of the 4D$_{\rm class}$ theory is that the
renormalization deliver it from the invariance under time reversal which
would contradict to the macroscopic experience.
It is significant that this excess invariance is not violated at the expense 
of non-invariant terms of the Lagrangian, since
their presence would impair the invariance of the 3D$_{\rm quant}$ theory.
The violation of the AFF equivalence in the case $D=3$ is adequate to the
physical reality.
It can be liken to the chiral anomaly making possible the
decay $\pi^0\to 2\gamma$ which is forbidden due to the excess symmetry of
the standard model.

In the case $D=2$, the ultraviolet situation is qualitatively the same as in 
the case $D=3$: 
Divergences of both $\Sigma$ and $\delta m$ are absorbed by the mass 
renormalization.
However, the renormalized  3D$_{\rm class}$ dynamics is reversible.
Indeed, $c_1\sim\log\Lambda$, therefore, $c_2\sim\Lambda^{-1}$.
This means that  the mass renormalization makes no finite mark in the 
three-momentum of the `dressed' particle governed by the second Newton law.
Thus the AFF equivalence is observed. 

In the case $D=1$, there are no ultraviolet divergences at all,
and the AFF equivalence cannot be violated for this reason. 

\section
{Indeterminism in a classical realm}\label
{special}
Ultraviolet divergences is not the only reason why the AFF dualism is not
an exact equivalence.
Another reason is that classical pictures are devoid of quantum coherence, in 
other words, the classical determinism cannot be reconciled 
with the quantum principle of superposition.
The notion of probability is incorporated in classical statistical 
mechanics quite artifically; it expresses the measure of our ignorance of 
deterministic pictures in detail, whereas the probability amplitude is a 
fundamental element of quantum theory.
The holographic mapping of a classical theory onto a quantum theory should 
suffer information loss, but the mechanism of this loss is obscure\footnote{
To circumvent this difficulty, 't Hooft  \cite{Hooft99} suggested that
``Since, at a local level, information in these (classical) states is not
preserved, the states combine into equivalence classes.
By construction then, the  information that distinguishes the different
equivalence classes is absolutely preserved.
Quantum states are equivalence classes.''}.
We will give a schematic reasoning demonstrating that this problem
is tractable in the case $D=1$, namely, a kind of indeterministic behavior of 
particles is possible in the two-dimensional classical realm.

Let us write the dynamical equations of the discussed 2D$_{\rm class}$ theory:
\begin{equation}
\partial_\lambda F^{\lambda\mu}(x)=2e \sum_{a=1}^2\int_{-\infty}^\infty d\tau_a\,
{v}^\mu_a(\tau_a)\,\delta^{(2)}\Bigl(x-z(\tau_a)\Bigr),
\label
{maxw}
\end{equation}                                           
\begin{equation}
m_a {\dot v}^{\mu}_a=e_a {v}_{\nu}^a F^{\mu\nu}(z_a).
\label
{newt}
\end{equation}                                           
They are exactly integrable.
From the form of solutions to these equations,
two striking features of the 2D$_{\rm class}$ realm follow.
First, there is no radiation of electromagnetic waves in this realm \cite{k9}
and hence there is no dissipation of energy.
All motions of particles are reversible.
Second, it is possible that several point particles merge into a single
aggregate, and then, after a lapse of some period, split into the initial 
objects.

For simplicity, we restrict our consideration to a system of two particles
with equal masses $m$ and charges $e$ (denoting $e^2/m=a$) which are to be
related to the centre-of-mass frame.
Let the particles be moving towards each other, and their total energy is such
that, at the instant of their meeting, their velocities are exactly zero.
Then there exists an exact solution to Eqs. (\ref{maxw}) and (\ref{newt})
describing two worldlines $z^\mu_1(\tau)$ and ${ z}^\mu_2(\tau)$ which
coalesce at the instant $\tau^\ast$ and separate at the instant
$\tau^{\ast\ast}=\tau^{\ast}+T$,
\begin{equation}
z^\mu_1(\tau)=\cases{a^{-1}\{{\sinh} a(\tau-\tau^\ast),1-{\cosh} a(\tau-
\tau^\ast)\},& $\tau<\tau^{\ast}$\cr
\{\tau-\tau^{\ast},0\}, &$\tau^{\ast}\le\tau<\tau^{\ast\ast}$\cr
a^{-1}\{aT+{\sinh} a(\tau-\tau^{\ast\ast}), 
{\cosh} a(\tau-\tau^{\ast\ast}) -1\},& $\tau\ge\tau^{\ast\ast}$\cr},
\label
{z}
\end{equation}                                            
\begin{equation}
z^\mu_2(\tau)=\cases{a^{-1}\{{\sinh} a(\tau-\tau^\ast),{\cosh} a(\tau-
\tau^\ast)-1\},& $\tau<\tau^{\ast}$\cr
z^\mu_1(\tau), & $\tau^{\ast}\le\tau<\tau^{\ast\ast}$\cr
a^{-1}\{aT+{\sinh} a(\tau-\tau^{\ast\ast}), 
1-{\cosh} a(\tau-\tau^{\ast\ast})\},& $\tau\ge\tau^{\ast\ast}$,\cr}
\label
{-z}
\end{equation}                                            
and the retarded field $F^{\mu\nu}$ expressed through  
$z_1^\mu(\tau)$ and ${ z}_2^\mu(\tau)$ as
\begin{equation}
F^{\mu\nu}(x)=e\,\sum_{a=1}^2\,\frac{1}{\rho_a}\,(R_a^{\mu} {v}_a^{\nu}-
R_a^{\nu} {v}_a^{\mu}).
\label
{F}
\end{equation} 
Here, $R^\mu_a\equiv x^{\mu}-z^{\mu}_{a\hskip0.3mm{\rm ret}}$ is an isotropic
vector drawn from the emitting point $z^{\mu}_{a\hskip0.3mm{\rm ret}}$ on 
$a$th worldline to the point of observation $x^\mu$, and 
$\rho_a=R_a\cdot {v}_a$
the invariant retarded distance between $x^\mu$ and  
$z^{\mu}_{a\hskip0.3mm{\rm ret}}$.                                            

The parameters $\tau^{\ast}$ and $\tau^{\ast\ast}$ are arbitrary. 
If $\tau^{\ast}$ and $\tau^{\ast\ast}$ are different and finite, then the 
curves (\ref{z}) and (\ref{-z}) correspond to 
the formation of an aggregate with finite life time.
For $\tau^{\ast\ast}\to \infty$, they describe the formation of a stable
aggregate  never decaying.
For $\tau^{\ast}\to -\infty$, we see decay at a finite instant of an
aggregate formed at the infinitely remote past. 
If $\tau^{\ast}\to -\infty$ and $\tau^{\ast\ast}\to\infty$, then the curves 
degenerate into a straight line corresponding to an absolutely
stable aggregate.
For $\tau^{\ast}=\tau^{\ast\ast}$, they describe an aggregate which exists
for a single instant.
Thus the solution to Eqs. (\ref{maxw}) and (\ref{newt}) with the given
Cauchy data is not unique.
Moreover, we have a continuum of solutions since the period of the merged state
can be any $T\ge 0$.
The decay occurs quite accidentally at any instant.
Note, however, that variants of evolution with such Cauchy data comprise null
set.  

In summary, a $D+1$-dimensional classical picture can be equivalent to the AFF
associated $D$-dimensional quantum picture if and only if $D=1$.
Indeed, for $D=1$, the theory is free of ultraviolet divergences.
Only in two-dimensional classical realm, radiation is absent, 
the problem of dissipation of energy does not arise, and 
the dynamics is reversible.
Besides, there are variants of evolutions rendering dynamics 
stochastic already on the classical level.
Since the measure of these variants in phase space is zero, they have no
effect on the fulfilment of the Liouville theorem, so that the variable 
$t$ is outside the quantum description where survives only 
$ix_0$ playing the role of the `Euclidean' time.
Taking into account that clusters of classical charged particles in the 
two-dimensional realm mimic classical strings, it becomes clear that
the behavior of  classical strings is coded in the behavior of quantum
point objects, and a holographic equivalence between the classical 
string field theory and the local quantum field theory should exist.

\section{Conclusion}
\label
{Conclusion}
The holographic principle offers an intriguing possibility to interweave
the classical, the quantum, and topological features in different
dimensions.
In the absence of gravity, this principle can be realized with the aid of the 
AFF dualism.
In view of this dualism, the origin of anomalies turns out to be twofold.
First, the quantization increases the dimension of a given classical realm by
unite (leaving its classicality intact); this fact by itself can damage some
symmetries.
Second, the formal AFF correspondence is not equivalence due to the 
availability of ultraviolet divergences and conflict between classical
determinism and quantum indeterminism.
The violation of the AFF equivalence contributes to the symmetry damage.
In particular, it is responsible for irreversibility of classical dynamics.
The only opportunity to circumvent this nonequivalence is to lower
dimension to $D=1$.
Then classical stringy realm is likely can be holographically projected onto 
realm filled with quantum point objects.

\vskip5mm
\noindent
{\large\bf Acknowledgments}
\vskip2mm
\noindent
This work was supported in part by 
the International Science and Technology Center
under the Project {KR-154}.


\begin{thebibliography}{99}
\bibitem
{Jackiw} 
For an account of anomalies in conservation laws, see, e. g., 
R. Jackiw,  ``Field theoretic investigations in current algebra'' and 
``Topological investigation of quantized gauge theories'' in  \cite{Treiman}, 
and also his lecture
``The unreasonable effectiveness of quantum field theory,'' hep-th/9602122. 

\bibitem
{Treiman} 
S. Treiman,
R. Jackiw, B. Zumino and E. Witten, {\it Current Algebra and Anomalies}
(Princeton U. P./World Scientific, Princeton NJ, Singapore, 1985).

\bibitem
{Hooft}
G. 't Hooft, ``Dimensional reduction in quantum gravity,''
in: {\it Salam\-festschrift}. Edited by A. Ali, J. Ellis and S. Randjbar-Daemi
(World Scientific, Singapore, 1993), pp. 284-296.  gr-qc/9310006.

\bibitem
{Susskind}
L. Susskind, ``The world as a hologram,'' J. Math. Phys. 
{\bf 36} (1995) 6377-6396. hep-th/9409089.

\bibitem
{Maldacena}
J. Maldacena, ``The large N limit of superconformal field theories and
supergravity,''
Adv. Theor. Phys. {\bf 2} (1998) 231-252;
E. Witten, ``Anti de Sitter space and  holography,'' Adv. Theor.
Math. Phys. {\bf 2} (1998) 253-291; S. S. Gubser, I. R. Klebanov and 
A. M. Polyakov, ``Gauge theory correlators from noncritical string theory,'' 
Phys. Lett. {\bf B428} (1998) 105-114.


\bibitem
{Alfaro}
V. de Alfaro, S. Fubini and G. Furlan, ``Gibbs average in the functional 
formulation of quantum field theory,''
Phys. Lett. {\bf 105B} (1981) 462-466.

\bibitem
{Namiki}
M. Namiki, {\it Stochastic Quantization} (Springer, Heidelberg, 1992).

\bibitem
{Gozzi}
E. Gozzi, ``The new functional approach to field theory by de Alfaro, Fubini 
and Furlan and its connection  to the Parisi-Wu stochastic quantization,''
Phys. Lett. {\bf 130B} (1983) 183-188.

\bibitem
{vandeBruck}
C. van de Bruck, ``On gravity, holography and the quantum,'' gr-qc/0001048.

\bibitem
{Hooft99}
G. 't Hooft, ``Quantum gravity as a dissipative deterministic system,''
Class. Quant. Grav. {\bf 16} (1999) 3263-3279. gr-qc/9903084.

\bibitem
{HenningsonSkenderis}
M. Henningson and K. Skenderis, ``The holographic Weyl anomaly,'' 
JHEP 9807 (1998) 023. hep-th/9806087.

\bibitem
{Coleman} 
S. Coleman and E. Weinberg, ``Radiative corrections as the origin of symmetry
breaking,''
Phys. Rev. D {\bf  7} (1973) 1888-1910.

\bibitem
{Salam}
A. Salam, ``Renormalized S-matrix for scalar electrodynamics,''
Phys. Rev. {\bf 86} (1952) 731-744.

\bibitem
{teit}
C. Teitelboim,
``Splitting of Maxwell tensor: Radiation reaction without advanced fields,'' 
Phys. Rev. D {\bf 1} (1970) 1572-1582.

\bibitem
{k9} 
B. P. Kosyakov, ``Exact solutions in classical electrodynamics and 
the Yang-Mills-Wong theory in even-dimensional space-times,''
Theor. Math. Phys. {\bf 199} (1999) 493-505.

\end{thebibliography}
\end{document}